# Where Does the Density Localize? Convergent Behavior for Global Hybrids, Range Separation, and DFT+U

Terry Z. H. Gani[1] and Heather J. Kulik[1,*]

[1]*Department of Chemical Engineering, Massachusetts Institute of Technology, Cambridge, MA 02139*

ABSTRACT: Approximate density functional theory (DFT) suffers from many-electron self-interaction error, otherwise known as delocalization error, that may be diagnosed and then corrected through elimination of the deviation from exact piecewise linear behavior between integer electron numbers. Although paths to correction of energetic delocalization error are well-established, the impact of these corrections on the electron density is less well-studied. Here, we compare the effect on density delocalization of DFT+U, global hybrid tuning, and range-separated hybrid tuning on a diverse test set of 32 transition metal complexes and observe the three methods to have qualitatively equivalent effects on the ground state density. Regardless of valence orbital diffuseness (i.e., from 2$p$ to 5$p$), ligand electronegativity (i.e., from Al to O), basis set (i.e., plane wave versus localized basis set), metal (i.e., Ti, Fe, Ni) and spin state, or tuning method, we consistently observe substantial charge loss at the metal and gain at ligand atoms (ca. 0.3-0.5 e or more). This charge loss at the metal is preferentially from the minority spin, leading to increasing magnetic moment as well. Using accurate wavefunction theory references, we observe that a minimum error in partial charges and magnetic moments occur at higher tuning parameters than typically employed to eliminate energetic delocalization error. These observations motivate the need to develop multi-faceted approximate-DFT error correction approaches that separately treat density delocalization and energetic errors in order to recover both correct density and magnetization properties.

## 1. Introduction

Presently available exchange-correlation (xc) approximations in DFT are plagued by both one- and many-electron self-interaction errors (SIE)[1-5], also referred to as delocalization error[6-8] (DE), which give rise to well-known problems in dissociation energies[2, 9-12], barrier heights[13], band gaps[14-15], and electron affinities[16-18]. Although referred to as a delocalization error, this error is rigorously defined in energetic terms, with more indirect connection to the density itself. In particular, it is known[19] that an exact energy functional should be piecewise linear with respect to fractional addition ($q$) or removal of charge:

$$E(q) = (1-q)E(N) + qE(N+1) , \qquad (1)$$

where E($N$) and E($N$+1) are the energies of $N$- and $N$+1-electron systems, respectively, and $q$ is varied between 0 ($N$ electrons) and 1 ($N$+1 electrons). Both semi-local (e.g., local density approximation, LDA or generalized gradient approximation, GGA) functionals and the formally self-interaction free Hartree-Fock (HF) theory produce a deviation from piecewise-linearity[20] in E($q$) with convex and concave behavior, respectively[1, 21], and lack the associated derivative discontinuity[22-25] of the exact functional.

The correction of this deviation from linearity and elimination of energetic DE is achieved by invoking Janak's[26] or Koopmans'[27] theorem to both identify[28] and then correct an xc functional's curvature from the difference in the $N$+1-electron highest occupied molecular orbital (HOMO) and the $N$-electron lowest unoccupied molecular orbital (LUMO) eigenvalues:

$$\left\langle \frac{\partial^2 E}{\partial q^2} \right\rangle = \varepsilon_{N+1}^{\text{HOMO}} - \varepsilon_N^{\text{LUMO}} . \qquad (2)$$

By tuning the range separation parameters[29-38] that divide short-range semi-local or hybrid xc forms from long-range HF exchange to minimize this curvature, tuned range-separated hybrids[39-

[45] frequently improve excited state[46] and some ground state[33] properties by improving energetic delocalization error.[1, 4, 7-8, 14, 35, 47-48] The simplicity of this approach has led to its increased use over orbital-dependent corrections that remove one-electron SIE[49-54]. Prior to the development of range-separated hybrids, global hybrids with varying mixtures of HF exchange[55] had also been widely employed to approximately correct SIE and are still a main ingredient in flexible tuning of range-separated hybrids (see Sec. 2). Some of us recently demonstrated[56] that the DFT+U method[57-58], widely employed for treating SIE in transition metal chemistry[59], will never worsen energetic DE and may also recover piecewise linearity with varying degrees of efficiency. We also established that the appropriate $U$ should be equal to a scaled value obtained from eqn. 2, rather than the self-consistent[60], linear-response[61-62] $U$ calculated at fixed electron number. This observation can also be understood in the context of the marked size-dependence[63-66] of the range-separated hybrid tuning strategy, absent in equivalent calculations of linear-response $U$, which highlights that correction of energetic DE is not a sufficient criterion for SIE removal as system size grows.

Despite the successes of range-separated hybrids, global hybrids, and DFT+U for eliminating energetic DE, the impact of any of these strategies on properties of the density is less well-established. There are clear improvements to densities in pathological cases where the density DE and energy DE are both a result of charge transfer error[7-8, 67]. Some of us recently observed[68] global HF exchange tuning to universally localize electron density away from the metal and onto ligand states, consistent with range-separated hybrid studies that show decreased dative bonding in both 3$d$ states of the diatomic molecule CuCl[33] or in a representative iron octahedral complex[69] and 4$f$ states in lanthanide complexes[70]. Within DFT+U, the potential that favors localization should enhance filling of $d$ or $f$ states that are more than ½ filled while emptying states that are less than ½ filled (see Sec. 2), and the connection in the solid state community of DFT+U to the Hubbard[71] and Anderson[72] model Hamiltonians frequently invokes a statement regarding electrons being localized onto a metal site, not away from it, suggesting DFT+U could behave differently from hybrid functionals. Correct densities are a necessity for interpreting trends in chemical bonding[73] and associated observable quantities. Some functionals well-known to provide good energetics (e.g., B3LYP[74-76]), however have been demonstrated to yield poor densities in comparison to accurate references[77], while others may yield poor energetics and good densities[78]. Replacement of approximate DFT densities with ones derived from HF have been demonstrated to yield improved barrier heights[79-80] and dissociation energies[6, 81-82], enabling a separation of energetic- and density-driven delocalization errors[6]. Thus, it becomes clear that continued advancement of xc approximations necessitates consideration of delocalization errors both energetic and density-oriented in nature.

In this work, we provide a comprehensive demonstration of the universal nature of density localization in transition metal complexes from metal to ligand across 32 complexes that span varying ligand diffuseness and electronegativity, varying metal occupations, and both stretched and compressed bonds, regardless of the delocalization error correction (i.e., DFT+U, global hybrids, or range separated hybrids) employed. We also explore the extent to which this energetic delocalization error correction reduces density errors with respect to accurate wavefunction theory references. The rest of this article is outlined as follows. In sections 2 and 3, we provide the Theoretical and Computational Details, respectively, of the methods and calculations employed in this work. In section 4, we present Results and Discussion of the nature of density localization. Finally, in section 5, we provide our Conclusions.

## 2. Theoretical Details

In this work, we compare three strategies for treating energy- and density-delocalization error.

*Global Hybrid Functionals*. Hybrid functionals are widely employed for approximately correcting self-interaction errors in practical DFT. Both low[83-85] and high[86-88] percentages of Hartree-Fock (HF) exchange have been proposed for the accurate description of transition metal complexes, likely due to the increasing static correlation error[4] with increasing HF exchange, as diagnosed by increasing fractional spin error[89]. The well-known B3LYP[74-76] hybrid is defined as:

$$E_{xc}^{B3LYP} = E_x^{LDA} + a_0(E_x^{HF} - E_x^{LDA}) + a_x(E_x^{GGA} - E_x^{LDA}) + E_c^{LDA} + a_c(E_c^{GGA} - E_c^{LDA}) \quad (3)$$

where $a_0$=0.20 (20% exchange), and the GGA (B88) enhancement factors over LDA are $a_x$=0.72 and $a_c$=0.81 for exchange and correlation, respectively. We again employ a modified[68] B3LYP exchange expression to enable HF exchange tuning:

$$E_x^{modB3LYP} = E_x^{LDA} + a_0(E_x^{HF} - E_x^{LDA}) + 0.9(1 - a_0)(E_x^{GGA} - E_x^{LDA}) \quad (4)$$

while holding the GGA/LDA ratio fixed to the 9:1 value in standard B3LYP[74-76].

*Range-Separated Hybrid Functionals*. The most widely-utilized[35] approach for correcting energetic delocalization error is to employ a range-separated[90] hybrid functional, which introduces a distance-dependent Coulomb repulsion operator:

$$\frac{1}{r_{12}} = \frac{1 - [\alpha + \beta \mathrm{erf}(\omega r_{12})]}{r_{12}} + \frac{\alpha + \beta \mathrm{erf}(\omega r_{12})}{r_{12}} , \quad (5)$$

where the first, short-range potential term decays on a $1/\omega$ length-scale, and the second term is a long-range potential with correct $1/r$ asymptotic behavior for $\alpha+\beta=1$. We restrict the focus in this work to introducing HF exchange in the long-range portion and employing semi-local DFT in the short range (i.e., $\alpha$=0, $\beta$=1), where eqn. 5 then takes the form:

$$\frac{1}{r_{12}} = \frac{1-\mathrm{erf}(\omega r_{12})}{r_{12}} + \frac{\mathrm{erf}(\omega r_{12})}{r_{12}} \quad , \tag{6}$$

specifically with the LRC-ωPBE functional[91], which uses PBE[92]-GGA in the short-range.

*DFT+U.* The full DFT+U energy functional[58, 93] may be expressed as:

$$E_{\mathrm{DFT+U}}[n(\mathbf{r})] = E_{\mathrm{DFT}}[n(\mathbf{r})] + E_{\mathrm{Hub}}[\{n_m^{I\sigma}\}] - E_{\mathrm{DC}}[\{n^{I\sigma}\}] \quad , \tag{7}$$

where the first term (DFT) is the contribution from any xc approximation, the second term (Hub) is a Hubbard model Hamiltonian correction, and the double counting (DC) term approximately removes the effect of corrections present in both of the first two terms. There will be a Hub and DC contribution for each Hubbard atom and subshell identified. By employing a DC term obtained within the fully-localized limit and making a frequent simplifying assumption[62, 94] to treat same-spin and opposite-spin electrons equivalently (i.e., $U_{\mathrm{eff}}$=$U$-$J$) we obtain an expression for the DFT+U energy as:

$$E^{\mathrm{DFT+U}} = E^{\mathrm{DFT}} + \frac{1}{2}\sum_{I,\sigma}\sum_{nl} U_{nl}^{I}[\mathrm{Tr}(\mathbf{n}_{nl}^{I\sigma}(\mathbf{1}-\mathbf{n}_{nl}^{I\sigma}))] \quad . \tag{8}$$

There is a "+U" contribution for each *nl* subshell of atom *I* to which a $U_{nl}^{I}$ is applied. The elements of the $\mathbf{n}_{nl}^{I\sigma}$ occupation matrix are obtained as a projection of the molecular state $|\psi_{k,v}\rangle$ at k-point *k* onto localized atomic orbitals on an atom *I*:

$$n_{mm'}^{I\sigma} = \sum_{k,v} \langle \psi_{k,v} | \phi_{m'}^{I} \rangle \langle \phi_m^{I} | \psi_{k,v} \rangle \quad . \tag{9}$$

The "+U" correction is incorporated self-consistently with a modification to the potential as:

$$V^U = \sum_{I,nl}\sum_{m}\frac{U_{nl}^I}{2}(1-2n_{nl,m}^{I\sigma})\left|\phi_{nl,m}^I\right\rangle\left\langle\phi_{nl,m}^I\right| . \tag{10}$$

The range of effects of the "+U" functional on both the total energies and molecular orbital energies for energetic delocalization error correction has been recently outlined by some of us.[56] The Hubbard U corresponds to the difference between the ionization potential (IP) and electron affinity (EA) of electrons on atom *I* in subshell *nl* with respect to the rest of the system, which is a finite difference approximation to the second derivative of the energy (i.e., $U_{nl}^I = \frac{\partial^2 E}{\partial (n_{nl}^I)^2}$) that may be calculated, as outlined in Ref. [59]. Comparison between DFT+U and hybrid methods has been of recent interest[95] due to the ease of use of the former in periodic boundary conditions and the latter in gas phase calculations.

## 3. Computational Details

*Global and range-separated exact exchange*. The effect of exact exchange was investigated by altering[68] the percentage of Hartree-Fock (HF) exchange in a modified form of the B3LYP[74-76] global hybrid functional from as low as 0% (i.e., a pure BLYP GGA) to as high as 40% HF exchange in increments of 10%, unless otherwise noted. The effect of long-range Hartree-Fock exchange was investigated by altering the range-separation parameter ω in the ωPBE range-separated hybrid functional[90], which mixes the pure Perdew-Burke-Ernzerhof (PBE)[92] GGA functional at short range with asymptotically correct HF exchange at long range. The value of ω varied in this work is from 0.0 bohr$^{-1}$ (i.e., pure PBE) to as high as 0.4 bohr$^{-1}$ in increments of 0.1 bohr$^{-1}$. Both sets of calculations were performed using the TeraChem[96-97] graphical processing unit (GPU)-accelerated quantum chemistry package with a localized basis set (LBS). The default definition of B3LYP in TeraChem employs the VWN1-RPA form for the LDA VWN[98] component of LYP[74] correlation. Ti, Fe, Ni, Se, and Te were treated with the LANL2DZ effective core potential[99-100], and the 6-31G* basis was used for the remaining atoms. All calculations were spin-unrestricted with virtual and open-shell orbitals level-shifted[101] by 1.0 and 0.1 eV, respectively, to aid self-consistent field (SCF) convergence to an unrestricted solution.

*DFT+U*. DFT+U calculations were performed using the periodic boundary condition code Quantum-ESPRESSO[102], which employs a plane-wave basis set (PWBS). The PBE[92] GGA was employed with ultrasoft pseudopotentials (USPPs)[103] obtained from the Quantum-ESPRESSO website[104]. Plane-wave cutoffs were 30 Ry for the wavefunction and 300 Ry for the charge density. A full list of USPPs used in this work is provided in Table S1 of the Supporting Information. The Martyna-Tuckerman scheme[105] was used in order to eliminate periodic image effects in the calculations on the molecular complexes studied. Cubic box dimensions ranging from 8 Å to 12 Å were employed depending on the size of the complex together with a 320 x 320 x 320 FFT grid, and a list of box dimensions for each complex is provided in Table S2 of the Supporting Information. Box sizes were chosen to ensure a reasonably small grid spacing for the accurate determination of Bader atomic charges. At least 15 and up to 25 unoccupied states (bands) were included for all complexes. The DFT+U correction on the 3$d$ states of the transition metal complexes employed projections onto atomic states obtained during pseudopotential generation, as is standard practice[59]. The Hubbard U values employed in this work ranged from 0 eV (i.e., pure PBE-GGA) to 5 eV in 1 eV increments, unless otherwise noted, following common practice in studies of tuning $U$ effects and typical ranges of $U$ values employed.[59, 106]

*Correlated wavefunction theory (WFT)*. Complete active space second-order perturbation theory (CASPT2)[107] calculations were performed with Molcas 8.0[108] on three representative geometries: equilibrium and stretched [Fe(H$_2$O)$_6$]$^{2+}$ and equilibrium [Fe(NH$_3$)$_6$]$^{2+}$. Calculations were carried out following the details described in ref [109]. Relativistic atomic natural orbital (ANO-rcc) basis sets[110-111] contracted to [7s6p5d3f2g1h] for Fe, [4s3p2d1f] for O and N and [3s1p] for H were used together with the scalar-relativistic Douglas-Kroll Hamiltonian[112-113]. An imaginary level shift of 0.1 and IPEA shift of 0.25 were also used[114-115]. Active spaces of 10 electrons in 12 orbitals were used for the equilibrium complexes but a smaller active space of 6 electrons in 10 orbitals was used for the stretched complex due to RASSCF convergence failure with the (10,12) active space.

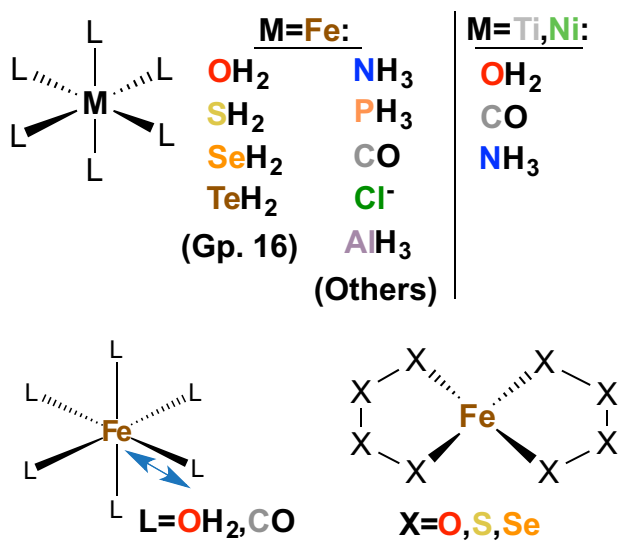

**Figure 1.** Summary of transition metal complexes studied in this work. (Top, left) Fe(II) octahedral complexes with group 16 ligands and ligands of varying field strength. (Top, right) Ti(II) and Ni(II) octahedral complexes of varying ligand field strength. (Bottom, left) Fe(II) octahedral complexes with weak-field water and strong field carbonyl ligands in compressed or stretched geometries. (Bottom, right) Distorted, square planar Fe(II) chalcogenide complexes with oxygen, sulfur, and selenium.

*Geometric structures and electron configurations.* A total of 32 different complexes (29 octahedral and 3 distorted tetrahedral) with varying metal centers and ligands were studied in this work (see Figure 1 for representative structures and Supporting Information Table S2 for a complete list). The polyselenide complex structure was obtained from the Cambridge Structural

Database[116] (accession code DIWLAY) through a search using the CCDC ConQuest web-screening tool limiting elements to Fe and Se. Analogous polyoxide and polysulfide structures were built by replacing the Se atoms with O and S atoms respectively. Structures for all other complexes were generated with the molSimplify[117] toolkit with the trained metal-ligand bond length feature enabled. These structures were subsequently geometry optimized with standard B3LYP in the charge corresponding to M(II) oxidation state and the ground high-spin state assigned to the isolated ion according to the National Institute of Standards and Technology atomic spectra database[118], i.e. quintet for Fe and triplet for Ti or Ni. This chosen spin state corresponds to the ground state for the ligands studied in this work, except for the strong-field ligand cases (e.g., CO) where low spin configurations may be preferred, but we compare effects within a fixed oxidation and spin state to simplify comparison, unless noted in the text.[56] The optimizations were carried out using the L-BFGS algorithm in translation and rotation internal coordinates[119] as implemented in a development version of TeraChem[96-97] to the default tolerances of $4.5 \times 10^{-4}$ hartree/bohr for the maximum gradient and $1 \times 10^{-6}$ hartree for the change in SCF energy between steps. Structures with non-equilibrium metal-ligand bond lengths were generated by direct bond length manipulation in Avogadro 2 0.8.0[12, 120] without subsequent constrained geometry optimization. Coordinates of all geometries studied in this work are provided in the Supporting Information.

*Partial charges and post-processing.* Natural population analysis (NPA)[121] partial charges and subshell occupations were obtained from the TeraChem interface with the Natural Bond Orbital (NBO) v6.0 package[122]. For quantification and comparison of partial charges across all three tuning procedures, we employ Bader atomic charges[73, 123], obtained from the BADER program[124], as they are solely functions of the spatial electron density (comparisons to alternate partial charge schemes are provided in the Supporting Information). The same grid resolution and dimensions were used across all methods for consistency. Vacuum charges not assigned to any atomic volume were ignored as the central position of the metal in all complexes ensures small and well-defined atomic volumes. Magnetic moments were calculated by integrating the spin density over the Bader atomic volumes for global and range-separated exact exchange, and by taking the difference of spin-up and spin-down Lowdin populations for DFT+U. Cube files and electron density differences were obtained using the Multiwfn post-processing package[125] for TeraChem calculations and the pp.x post-processing code for Quantum-ESPRESSO calculations.

## 4. Results and Discussion

### 4a. Charge Localization in Representative Complexes

We first consider the prototypical $[Fe(H_2O)_6]^{2+}$ complex which has been widely employed in previous studies on evaluating exchange-correlation functional choice[68, 86, 126], the role of exact exchange[68, 86], and curvature corrections[56] in transition-metal complexes. In this complex, a quintet spin-state Fe is in the +2 oxidation state with weak σ-donation from the ligands to the metal center, and the B3LYP-optimized structure employed in our analysis is a slightly distorted octahedron with average Fe-O bond length of 2.14 Å (see Supporting Information for coordinates).

We note that slight differences are present in the three GGA reference points we use for evaluating functional tuning strategies, namely: the PBE-GGA limit is achieved for ω-tuning and DFT+U but in a localized basis set (LBS) and plane-wave basis set (PWBS) formalism, respectively, whereas the HF exchange tuning $a_{HF}=0$ limit corresponds to a pure BLYP-GGA/LBS result. Nevertheless, differences for the GGA reference obtained from each of the three tuning strategies are slight, with all three methods providing a partial charge close to the overall average GGA Fe Bader partial charge of approximately 1.5 (see the left-most point in each plot in Figure 2). Regardless of tuning method employed, we observe that the already positive Fe partial charge uniformly increases with the tuning parameter for HF exchange, "+U" correction, and range-separation, signifying further charge localization away from the metal and onto the ligands. It may be expected that introduction of global and range-separated HF exchange impact the electron density equivalently. It is, however, surprising that the functional form of DFT+U produces the same effect since the "+U" potential may be expected to increase occupation of all half-filled atomic orbitals and only decrease occupation of less than half-filled atomic orbitals.

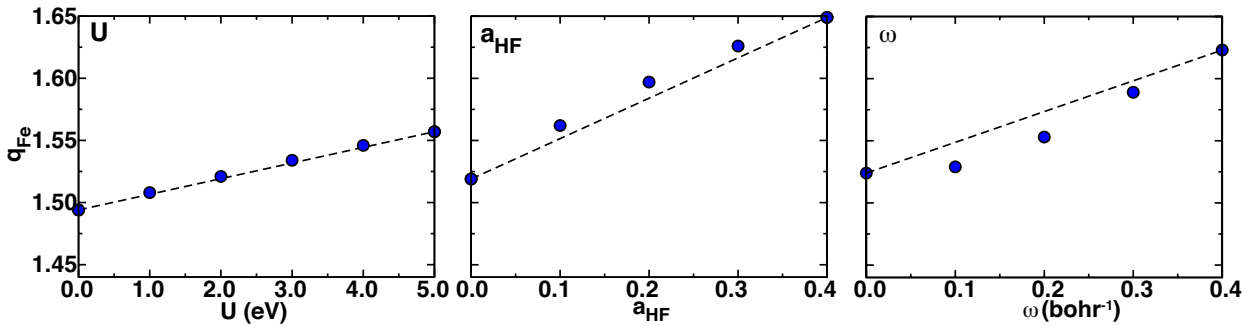

**Figure 2**. Dependence of Fe partial charge with $U$ (left), $a_{HF}$ (center), and $\omega$ (right) for the octahedral $[Fe(H_2O)_6]^{2+}$ complex. The black dashed lines indicate the linear approximation from displayed endpoints used to quantify the partial charge sensitivities, as described in the main text.

Whereas approximately linear behavior is observed across the range of $U$ and $a_{HF}$ studied (Figure 2, left and center), there is a marked nonlinear dependence on $\omega$ (Figure 2, right), which can be rationalized in terms of the functional form of the range-separated hybrid functional (see Section 2). We thus introduce linear fits as approximations to the Fe partial charge sensitivities ($S$), or the partial derivatives of the Fe partial charge with respect to each of the tuning parameters, $p$,

$$\text{slope} = S_p = \frac{\Delta q_{Fe}}{\Delta p} \approx \frac{\partial q_{Fe}}{\partial p}, p \in \{a_{HF}, U, \omega\}, \tag{11}$$

noting that the derivative with respect to ω is approximately evaluated from the endpoints at ω = 0.0 to 0.4 bohr$^{-1}$ to maintain consistency with the other tuning parameters. The units for $U$ and ω are eV and bohr$^{-1}$ respectively, and we use the unit notation "HFX"[68] to represent the range from 0% to 100% HF exchange. Some consideration should be made to the relative range of tuning typically employed for each of these three strategies to approximately correct SIE. Namely, the commonly proposed values of HF exchange for transition metal complexes in the literature range from around 0%[83-85] to at most 40-50%[86-88] (i.e., 0.5 HFX), and most optimally-tuned RSH results on analogous 4$f$ complexes[70] span a range of 0.0 to 0.5 bohr$^{-1}$ for ω. In comparison, typical ranges of $U$ would be from 0 to around 5 eV or more.[56, 59] Returning to the sensitivities for the hexa-aqua complex (Figure 2 and Supporting Information Table S4), we observe roughly 0.05 e loss from Fe with DFT+U tuning up to 5 eV, intermediate 0.10 e loss for ω up to 0.4, and 0.12 e loss for HF exchange up to 40%, indicating relatively comparable effects of these tuning parameters (see also Sec. 4e). Therefore, when reporting sensitivities we take into account the inequivalence of the order of magnitude of units in the denominator by multiplying all DFT+U sensitivities by 10, as indicated throughout the rest of this work (i.e., for Fe(H$_2$O)$_6^{2+}$, $S_p$ values are 0.33 e/HFX, 0.13 e/10 eV of $U$, and 0.25 e/bohr$^{-1}$, see Supporting Information Table S4).

Comparing our GGA-calculated and tuned Fe Bader partial charges against those calculated at the CASPT2 level of theory, we find fortuitously good agreement between the GGA and CASPT2 Fe partial charges of 1.5 for both methods in the hexa-aqua complex. Thus, increased electron density localization onto ligands by any of the three tuning strategies will worsen agreement with CASPT2 results. The GGA agreement is poorer, however, for the structurally and chemically similar hexa-ammine complex, with GGA and CASPT2 Fe partial charges of 1.4 and 1.6, respectively, suggesting a beneficial effect of approximate SIE corrections. Introducing and extrapolating a linear fit from eq. (11), approximately 65% HF exchange, $\omega=0.8$ Bohr$^{-1}$, or a $U$ of 24 eV, will bring the DFT partial charges for the hexa-ammine complex into quantitative agreement with those from CASPT2 (see Supporting Information Table S3 for details). We will return to these two complexes later in the text to identify whether DFT and WFT density properties obtained at functional parameters tuned to recover piecewise linearity are in improved agreement (see Sec. 4e).

To further investigate the specific origin of the observed density localization away from the metal center, we obtain the Fe 3$d$ and 4$s$ subshell occupancies as computed by the NAO scheme for HF exchange and LRC and an orthogonalized atomic projection scheme for the "+U" correction. Total occupancies of the t$_{2g}$ ($d_{xy}$, $d_{xz}$ and $d_{yz}$) and e$_g$ ($d_{x2-y2}$ and $d_{z2}$) AOs provide a good representation of the occupied electron density, whereas the 4$s$ AO may be neglected due to its occupancy being solely derived from hybridization with the 3$d$ AOs (i.e., Fe(II) has a nominal $d^6s^0$ electron configuration[118]). We observe that both the t$_{2g}$ and e$_g$ occupancies decrease as each tuning parameter increases, signifying a common physical origin of ligand charge localization across all three tuning procedures that is independent of orbital energy or character.

The extent of charge localization away from the metal center may also be visualized directly in terms of the electron density difference (Figure 3), which is defined as

$$\Delta\rho_p(\mathbf{r}) = -\int \frac{\partial \rho(\mathbf{r})}{\partial p} dp, p \in \{a_{HF}, U, \omega\}, \qquad (12)$$

where the limits of integration are discussed in Sec. 3. To ensure a fair comparison of the electron density difference across all three tuning procedures, we select an isovalue ($|\Delta\rho_c|$) of 0.002 e⁻/Å³ for $a_{HF}$ and normalize the isovalues for $U$ and $\omega$ according to the overall change in partial charge on the metal:

$$\frac{|\Delta\rho_c(p)|}{|\Delta\rho_c(a_{HF})|} = \frac{\Delta q_{Fe}(p)}{\Delta q_{Fe}(a_{HF})}, p \in \{U, \omega\} . \tag{13}$$

Across all three methods, we observe loss of electron density from the area directly surrounding the Fe center and gain of electron density in the areas surrounding the O atoms, corroborating our initial observations of increasing Fe partial charge and decreasing Fe 3$d$ AO occupancies. Slight differences nevertheless arise in how the additional electron density is distributed within the H$_2$O ligands. Increasing $a_{HF}$ and $\omega$ (Figure 3, center and right) results in a spatially uniform increase in the electron density around the O atoms with concomitant delocalization of charge away from the H atoms, whereas increasing $U$ (Figure 3, left) appears to localize the additional electron density to the O atom and does not affect the electron density around the H atoms. This distinct behavior in DFT+U should be expected, as the "+U" potential shift applied only affects molecular orbitals that contain 3$d$ atomic orbital character[56] (here, Fe-centered or Fe-O hybridized states). Thus, on preliminary test cases, all three strategies remove density delocalization error from approximate DFT by localizing density onto ligand atoms, consistent with previous results obtained on HF exchange[68]. Nevertheless, further examples will establish the universality of this observation.

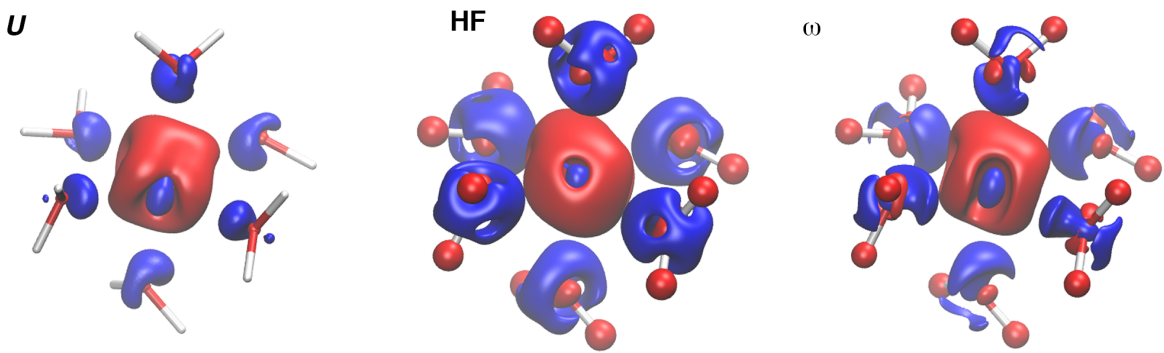

**Figure 3**. Isosurfaces of the *U* (left), global HF exchange (center), and ω (right) charge difference for the octahedral $[Fe(H_2O)_6]^{2+}$ complex, with geometric structure shown as sticks. The isovalues employed (see main text for derivation) are 0.00097 e$^-$/Å$^3$ (left), 0.002 e$^-$/Å$^3$ (center), and 0.0015 e$^-$/Å$^3$ (right). Red and blue volumes represent regions of negative (electron density lost) and positive (electron density gained) electron density difference, respectively.

**4b. Ligand Diffuseness and Electronegativity Effects**

A commonly-invoked depiction of self-interaction error (SIE) in transition-metal complexes is derived from the argument that 3*d* electrons are well-localized and thus subject to larger magnitudes of SIE than extended states, e.g. in bulk metals or in delocalized covalent bonding in molecular systems. One conceivable argument for why charge transfer is observed instead from the metal to oxygen ligands in the previously described hexa-aqua system (Sec. 4a) is that the 2*p*- and 1*s*-derived molecular orbitals of water are much more well-localized than the 3*d* states of the central iron atom. This relative localization of the two constituents is evident from the characteristic length, defined as the largest distance from the atomic center to the 0.001 *e* isosurface, of 2*p* and 3*d* orbitals for O and Fe atoms, of 1.7 Å and 1.9 Å, respectively (Figure 4). Such an observation motivates comparison to ligands with orbitals substantially more diffuse than 3*d* states of the iron atom, creating scenarios where the metal valence states are truly well-localized with respect to surrounding ligands. If the effect of relative diffuseness dominates, we may expect the direction of charge localization with approximate SIE correction (i.e., DFT+U and global or range-separated hybrid tuning) to reverse.

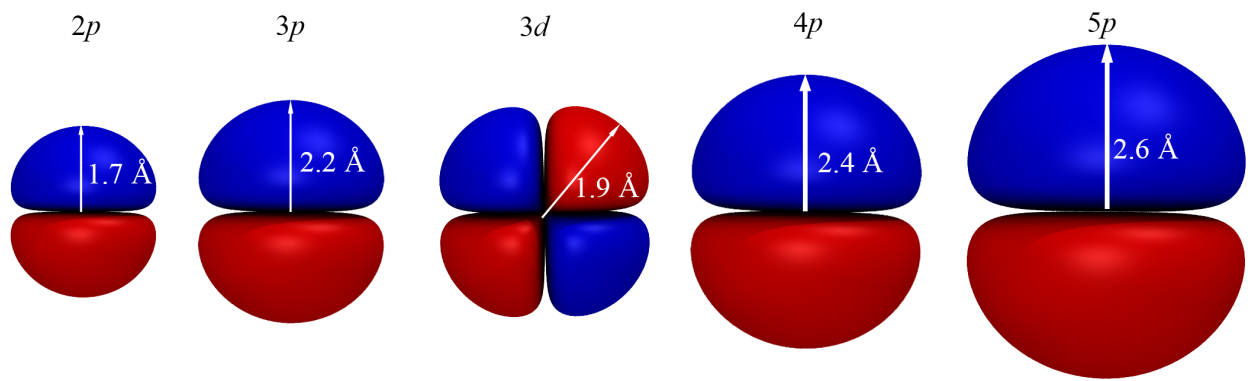

**Figure 4.** Schematic of increasing $nl$ subshell atomic orbital size from left to right: $2p$, $3p$, $3d$, $4p$, and $5p$ with annotated characteristic length and direction from the center of the atom to the furthest point on the 0.001 $e^-$ isosurface, as described in the main text.

Thus, we compare the iron hexa-aqua complex with the series of octahedral $[Fe(H_2X)_6]^{2+}$ complexes where X=S, Se, and Te corresponding to increasingly heavy group 16 elements with valence $3p$, $4p$ and $5p$ orbitals, respectively. We again quantify the increasing diffuseness of the underlying AOs of each of these ligands in terms of the characteristic length, which is 2.2 Å for $3p$ orbitals in S, 2.4 Å for $4p$ in Se, and 2.6 Å for $5p$ in Te, all larger than the 1.9 Å $3d$ orbitals of Fe (Figure 4). This series of weak field ligands gives structurally stable complexes in DFT gas phase geometry optimizations, despite only $[Fe(H_2O)_6]^{2+}$ being stable experimentally[127] due to increasingly weak Fe-X bonds for heavier substituents. We first note that the Fe partial charge computed at the GGA level of theory (Figure 5) decreases (i.e., becomes more neutral) as we proceed down the group, consistent with the trend in ligand electronegativity χ (Pauling scale – O: 3.4 > S: 2.6 > Se: 2.6 > Te: 2.1). This increasingly favorable, charge localization onto the metal may alternatively be interpreted through increased energetic overlap between the Fe $3d$ states and surrounding ligand states (Supporting Information Figure S1). Thus, within a constant functional choice, relative localization of the $3d$ states with respect to surrounding atoms does encourage 3d delocalization to the ligands.

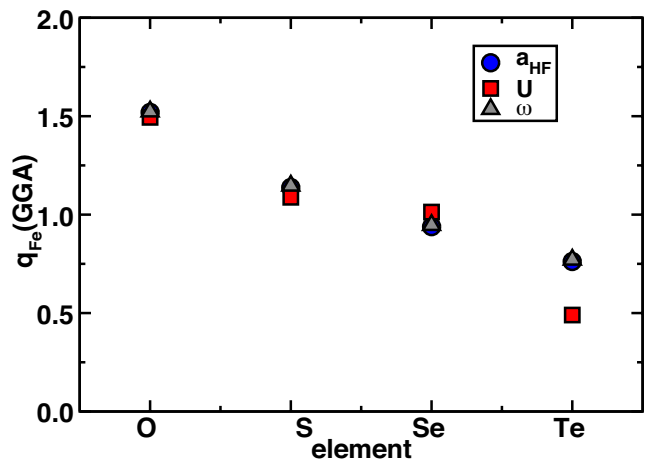

**Figure 5**. Fe GGA reference partial charge for HF exchange (BLYP/LBS, blue circles), $U$ (PBE/PWBS, red squares) and $\omega$ (PBE/LBS, gray triangles) for octahedral $[Fe(H_2X)_6]^{2+}$ complexes where X = O, S, Se, Te.

Sensitivity ($S_p$) of the partial charge on iron with respect to functional tuning parameter, $p$, (i.e., $U$ in DFT+U, $\omega$ in range correction, or $a_{HF}$ in global HF exchange), however, likely provides a more useful guide to interpret electron delocalization with respect to the GGA reference. Focusing on the direction of charge localization with functional tuning for each complex, we observe that all tuning approaches applied to all substituent ligands lead to increased charge localization away from the metal (Figure 6). For increasingly diffuse ligand orbitals, the positive sensitivities do not change sign and instead increase in magnitude down the group for all tuning methods (Figure 6). The largest increase observed for $\omega$-tuning from 0.25 e/bohr$^{-1}$ for O to 0.64 e/bohr$^{-1}$ for Te may be rationalized by the additive effect of increased bond length (2.1 Å for Fe-O vs. 3.1 Å for Fe-Te) that does not affect the more modest increases in global hybrid tuning sensitivities from 0.33 e/HFX to 0.67 e/HFX (Figure 6 and Supporting Information Table S4). The DFT+U sensitivities are even less element-sensitive, increasing only to 0.23 e/10 eV of $U$ for Te vs. 0.13 e/10 eV of $U$ for O. Overall, there is an apparent qualitative inverse correlation between the GGA partial charge and $S_p$, but the trend across the full data set employed in this work is very weak ($R^2$=0.08-0.14, see Supporting Information Table S4 and Figure S2).

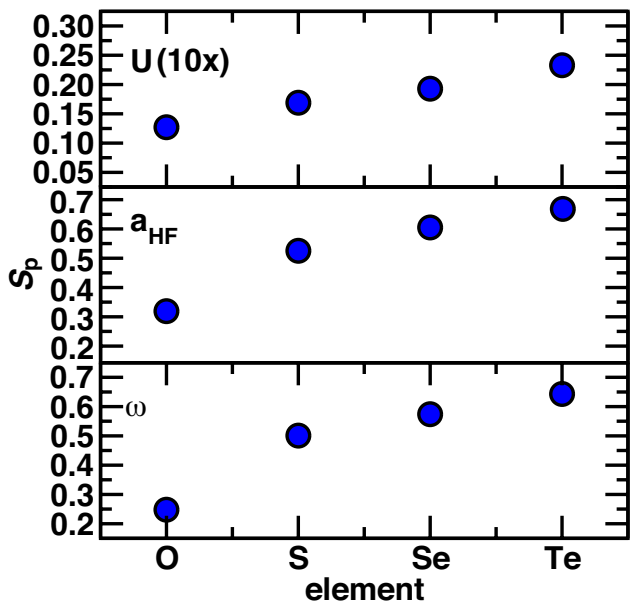

**Figure 6.** Fe partial charge sensitivity to changes in $U$ (e/eV x10, top), HF exchange (e/HFX, middle), and ω (e/bohr$^{-1}$, bottom) for the octahedral [Fe(H$_2$X)$_6$]$^{2+}$ complexes where X = O, S, Se, Te. The y-axes are the same for the global and range separated hybrid variations, whereas the DFT+U sensitivities have been multiplied by 10 due to differences in units and span a range half as large.

Finally, to verify that the observed effects are specific only to the diffuseness of the bonding ligand atom orbitals and hence generalizable to other complexes not pursued in this work, we also included negatively charged four-coordinate polychalcogenide complexes ([Fe(X$_4$)$_2$]$^{2-}$, X = O, S, Se) in our data set (structures shown in Figure 1). The qualitative partial charge (i.e., O > S > Se) and parameter sensitivity (i.e., O < S < Se) orderings are obeyed; quantitatively partial charges are lower on iron and sensitivities are about 10% higher overall but show less elemental-dependence (Supporting Information Table S4). The reduced dependence on element identity can likely be traced to 0.1-0.2 Å shorter bond distances in the polychalcogenide complexes (e.g., Fe-Se of 2.6 Å vs. 2.8 Å in the octahedral complex) that correspond to stronger coordination even for heavier elements (Supporting Information Table S2). Thus, we can motivate the extension of our observations of charge localization onto ligand atoms beyond octahedral complexes to other coordination environments.

As we have attributed differences in absolute partial charges at the GGA level of theory to differences in ligand electronegativity, it may be proposed that the positive sign of $S_p$ values can be partially attributed to positive ligand-metal electronegativity differences among the complexes considered above ($\chi$=1.8 for Fe). To assess the validity of this hypothesis, we constructed a somewhat artificial high-spin octahedral $[Fe(AlH_3)_6]^{2+}$ complex in which the direct ligand atom (Al, $\chi$=1.6) electronegativity is less than the iron electronegativity. As this complex dissociates upon gas phase geometry optimization, we modify the optimized geometry of $[Fe(NH_3)_6]^{2+}$ by assigning bond lengths equal to the sum of covalent radii[128] (2.7 Å for Fe-Al and 1.5 Å for Al-H). Consistent with the negative ligand-metal electronegativity difference and electron-deficient nature of $AlH_3$, we indeed observe a negative Fe partial charge of -1.0 at the GGA level of theory. Positive $S_p$ values of 0.16 e/10 eV of $U$, 0.12 e/HFX, and 0.26 e/bohr$^{-1}$, which correspond to the negative Fe partial charge becoming more neutral and charge localizing onto the $AlH_3$ ligands, are obtained for DFT+U, global- and range-separated-hybrid tuning, respectively (see Supporting Information Table S4). Compared to a hexa-aqua complex reference (see Sec. 4a), these sensitivities are reduced for the global-hybrid tuning, but they are comparable or larger for the DFT+U or range-separated hybrid methods. Thus, we confirm that the delocalization of charge from metal to ligand with three diverse functional tuning strategies is observed regardless of ligand orbital diffuseness or substituent atom electronegativity.

**4c. Non-equilibrium Complexes**

Another manifestation of approximate DFT failures is in the unphysical delocalization of charge near the dissociation limit[9] (e.g., in NaCl[9, 82], $CH^{+82}$ and $CO^{-1}$), resulting in spurious fractional charges, which may be attributed to either self-interaction error or to static-correlation error. SIE correction schemes such as the Perdew-Zunger approach[49] have been noted to eliminate such errors at dissociation at the cost of worsening equilibrium bond-length and density-derived properties[9]. It is also known that approximate SIE corrections, such as incorporation of an admixture of HF exchange, may increase static correlation error (SCE), owing to the higher SCE in HF, as quantified through fractional spin error[4]. Thus, the interplay of SIE and SCE motivates our examination of test cases with metal-ligand bond distances displaced from equilibrium values. In addition to the weak-field, hexa-aqua system (Sec. 4a), we consider the $[Fe(CO)_6]^{2+}$ complex with strong-field π-acceptor ligands as a case with contrasting metal-ligand bonding.

As we increase the bond length in both complexes away from equilibrium values of 2.1 and 2.3 Å for the Fe-O and Fe-C bonds, respectively, the Fe partial charge computed at the GGA level of theory passes through a maximum value at around $r = 1.2r_e$ before starting to decrease, with this effect slightly reduced for the PWBS partial charges with respect to the LBS GGA values (Figure 7). In the absence of SIE, we would expect the metal partial positive charge to increase with increasing bond length due to decreasing ligand-metal electron donation. This expectation is confirmed by CASPT2 $q$(Fe) partial charges of 2.0 in a stretched water complex with $r = 1.5r_e$, i.e. full charge delocalization away from the less electronegative Fe center. If the effect is SIE-dominant due to heterolytic dissociation, we may expect sensitivity for all three methods to increase with increasing bond length. Nevertheless, we may also anticipate the divergence in the three methods for SIE-correction to become more apparent for stretched bonds, as DFT+U is inherently short-range, acting only on metal states regardless of the placement of coordinating atoms, whereas the nature of range-separation means that separated ion interactions are increasingly treated with HF exchange as the atoms dissociate.

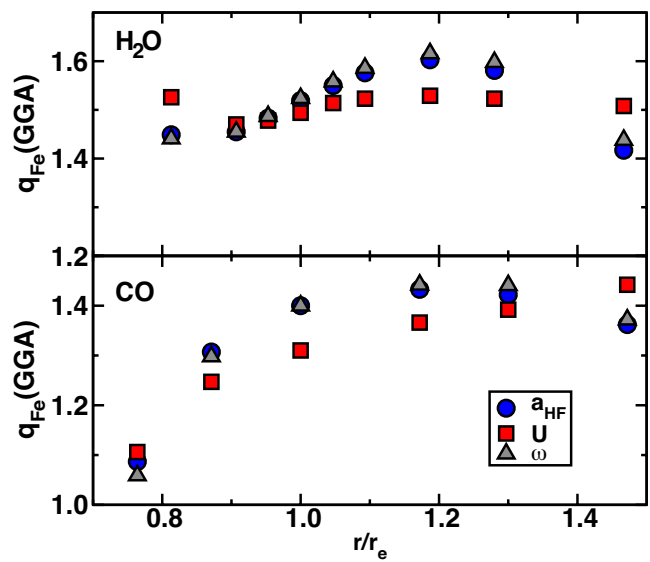

**Figure 7**. Dependence of Fe reference partial charge for HF exchange (BLYP/LBS, blue circles), $U$ (PBE/PWBS, red squares) and $\omega$ (PBE/LBS, gray triangles) with relative bond length ($r/r_e$) for $[Fe(H_2O)_6]^{2+}$ (top) and $[Fe(CO)_6]^{2+}$ (bottom), with respect to the equilibrium bond length ($r_e$) computed for the B3LYP equilibrium geometry.

If these functional tuning procedures are efficient mitigators of SIE, we thus expect $S_p$ to not only remain positive but increase as bonds are stretched away from equilibrium values, further extending what is encompassed by the near-universality of ligand charge localization as DFT+U and global- or range-corrected hybrid methods are applied. Indeed, we observe increasingly positive charge-sensitivities in both the hexa-aqua and hexa-carbonyl complexes as bonds are stretched by up to 50% beyond equilibrium values (Figure 8). All three methods exhibit comparable positive bond-length dependence of parameter sensitivity within $1.2r_e$ ($1.3r_e$) for the hexa-aqua (hexa-carbonyl) complex. For longer bond lengths, the three methods show divergent behavior, with the sensitivity of range-separation increasing dramatically at $1.5r_e$ to as much as $4S_p(r_e)$ (i.e., 0.25 e/au$^{-1}$ at equilibrium vs. 1.03-1.05 e/au$^{-1}$ for either complex). DFT+U sensitivities demonstrate the least geometric dependence, at around $2$-$2.5S_p(r_e)$ (i.e., 0.07-0.1 e/10 eV of $U$ vs. 0.2 e/10 eV of $U$), which can be rationalized as a combination of i) greater flexibility for electron delocalization in the PWBS and ii) the metal-centered nature of the DFT+U correction applied. Direct inclusion of intersite terms in DFT+U[129-130] to address metal-ligand bonding could produce more comparable behavior to hybrid functionals. Global-hybrid exchange demonstrates intermediate sensitivity increases at around $3S_p(r_e)$, (i.e., 0.29-0.33 e/HFX vs. 0.94-1.04 e/HFX, sensitivities for all complexes are provided in Supporting Information Table S4). Upon bond compression, metal charges generally become more negative and sensitivities are conversely reduced (see Figure 8 and Supporting Information Text S1).

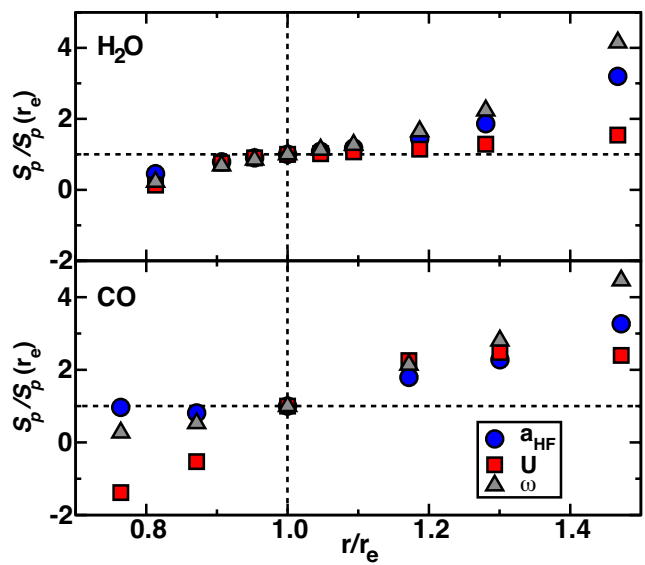

**Figure 8**. Dependence of relative Fe partial charge sensitivity ($S_p/S_p(r_e)$) for HF exchange (blue circles), $U$ (red squares), and $\omega$ (gray triangles) with relative bond length (r/r$_e$) for [Fe(H$_2$O)$_6$]$^{2+}$ (top) and [Fe(CO)$_6$]$^{2+}$ (bottom). Dashed black lines indicate the equilibrium sensitivity and bond lengths (i.e., $S_p/S_p(r_e) = 1$ and r/r$_e = 1$). The equilibrium bond length used in all cases is obtained from the B3LYP equilibrium geometries.

The observation of increasing sensitivity with bond length for global hybrids is consistent with previous observations of increasing differences in partial charges at the GGA and HF levels of theory with bond length in heterolytic dissociation of NaCl[9] and CH[+82]. Hence, these observations of both electronegativity- and ligand-derived localization are likely transferable beyond the transition metal complexes studied in this work. Nevertheless, despite increasing sensitivity with increasing bond length, the methods employed would have to be tuned beyond typically-applied values to recover CASPT2 partial charges on the iron in the stretched bond case. Our results also highlight the relative impact of approximate SIE corrections on the overall electron density: global and range-separated corrections within the realm of those under consideration may change partial charges on metal centers by as much as 0.6 e, leading to a substantial difference in SIE-derived delocalization, but DFT+U sensitivities correlate to more modest changes in the density over values of $U$ typically employed (i.e., 5-10 eV).

**4d. Effect of Metal Electron Configuration**

Having determined the effects of ligand identity and bond length on charge delocalization in a wide and representative array of Fe(II) complexes, we now turn our attention to the electron configuration of the central metal atom by examining paradigmatic early (Ti(II), $d^2$) and late (Ni(II), $d^8$) octahedral transition metal complexes with both strong-field CO and weak-field $NH_3$ or $H_2O$ ligands (structures shown in Figure 1). Again, positive partial charges and sensitivities, $S_p$, are observed for these complexes with one exception, $[Ti(CO)_6]^{2+}$, where the sensitivity is instead found to be weakly negative due to the partial charge scheme employed (see Supporting Information Text S2).

One might expect the Fe(II) complexes to be maximally sensitive to charge redistribution due to the half-filled character of the $3d$ states. Instead, we observe that $S_p$ for global- and range-separated hybrid tuning increases with $d$-electron count (i.e., ω-tuning and HF exchange $S_p$ values of ca. 0.20 e/bohr$^{-1}$ and 0.25 e/HFX for Ti < 0.25 e/bohr$^{-1}$ and 0.29-0.38 e/HFX for Fe <

0.3-0.4 e/bohr$^{-1}$ and 0.42-0.55 e/HFX for Ni, respectively) and shows greater metal $d$-filling-sensitivity than ligand-strength-sensitivity (Figure 9). Although this trend of doubling charge sensitivities from Ti to Ni is clearest with ω-tuning, unlike the group 16 elements, there is no additive size effect here, as all of the octahedral complexes considered have comparable metal-ligand bond lengths in the range of 2.1-2.3 Å. In contrast to the hybrid tuning strategies, DFT+U shows more limited metal-dependence, with comparable sensitivities in the range of 0.12-0.15 e/10 eV of $U$, except for carbonyl complexes that exhibit reduced sensitivities. Despite strong metal-dependent effects, subtler ligand-derived trends in previously observed for Fe(II) (CO < $H_2O$ < $NH_3$) are largely preserved for Ni(II) and Ti(II) complexes (Figure 9). Thus, the positive sign for sensitivities across the periodic table suggests that general trends we have observed for Fe(II) complexes likely hold for the remainder of the first-row transition metals.

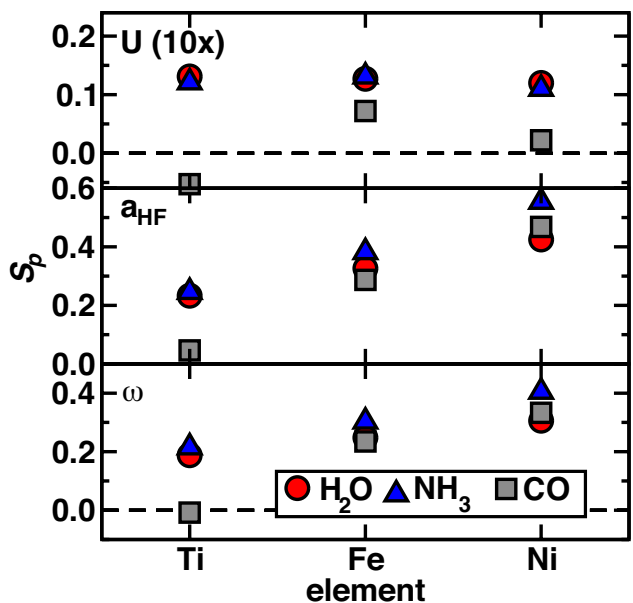

**Figure 9.** Ti, Fe, and Ni partial charge sensitivity to changes in $U$ (e/eV x10, top), HF exchange (e/HFX, middle), and ω (e/bohr$^{-1}$, bottom) for octahedral $[M(L)_6]^{2+}$ complexes where L = $H_2O$, $NH_3$, or CO. The y-axis ranges are the same for the global and range-separated hybrid variations, whereas the DFT+U sensitivities have been multiplied by 10 due to differences in units and span a range half as large.

Previously, some of us observed[68] positive $S_p$ for HF exchange across a series of Fe(II) and Fe(III) low-spin (LS) and high-spin (HS) complexes, suggesting that charge delocalization away from the metal is generalizable across oxidation and spin states as well. Based on the findings in the present work, we may expect similar trends for DFT+U and range-separation. In the same study[68], sensitivities were larger in the LS state than the HS state (i.e., the LS state loses charge faster than the HS state) across the majority of Fe(II) and Fe(III) complexes studied. Considering that the HS complexes had significantly longer metal-ligand bond lengths than the LS complexes, this result contradicts expectations based solely on geometry for a fixed electronic state (as previously illustrated in Figure 8), suggesting the importance of the metal's electron configuration. We thus expanded our molecule set to include a LS hexa-aqua Fe(II) complex computed at the HS equilibrium geometry. We indeed obtain low-spin Sp values roughly 50% higher than high-spin values consistently for all three tuning methods (see Supporting Information Table S4). Such an observation can be rationalized now in the context of our observations on increasing sensitivity with increasing $d$ filling. Rather than filling alone, it is the number of bonding orbitals occupied in complexes (i.e., the main distinguishing feature between HS and LS complexes for a given metal and oxidation state) that drives increased sensitivities. Although no one heuristic (e.g., ligand field strength or electronegativity) explains all trends in charge redistribution with functional tuning, all of these cases together serve to highlight the universality of charge localization toward ligands and away from the metal in transition metal complexes regardless of SIE-correction strategy.

**4e. Curvature Energy Corrections and Density or Magnetization Effects**

In addition to reproducing benchmarks or experimental values, functional-tuning strategies are increasingly employed to reproduce piecewise linearity, as indicated by highest-occupied and lowest-unoccupied molecular orbital (HOMO and LUMO) energies that correspond to total energy IPs and EAs, respectively. This approach is widely-employed in the context of optimally-tuned range-separated hybrids[35], but some of us recently demonstrated that DFT+U may also recover piecewise linearity[56]. We now quantify the relative extent to which the integer electron endpoint charge density is impacted by DFT+U, range-separation, or global hybrid exchange tuning that recovers piecewise linearity between integer-electron endpoints. We apply a simplified tuning procedure for all three tuning parameters in which the discrepancy between the LUMO and total-energy electron affinity of M(III) complexes is set to zero. For our M(II) complexes, we can thus write:

$$\text{LUMO error} = \varepsilon_{M^{3+}}^{\text{LUMO}} - E(M^{2+}) + E(M^{3+}) \;, \tag{14}$$

where the energies and eigenvalues are evaluated at the M(II) optimized geometries. The motivation for employing LUMO error alone to identify tuning parameters is two-fold: i) HOMO and LUMO errors are typically comparable magnitude and corrected with the same efficiency, otherwise necessitating a variable-parameter approach to recover piecewise linearity[56, 131] and ii) DFT+U tuning efficiencies are more sensitive to the LUMO than the HOMO[56]. To simplify the notation throughout, we use $p^*$ to represent the value of tuning parameter $p \in \{a_{HF}, U, \omega\}$ projected to eliminate the LUMO error based on LUMO error tuning trends extrapolated from the parameter ranges studied throughout this work (see Sec. 3 and Supporting Information Figure S4). In doing so, we aim to answer a few outstanding questions regarding the approximate correction of SIE: i) although piecewise linearity is essential for correct orbital energies, how does its recovery also impact integer-electron count electron densities and magnetizations with respect to accurate WFT references? ii) does recovering piecewise linearity require comparable tuning parameters between differing methods? iii) what is the overall relative effect of each tuning method on the ground state density upon recovery of piecewise linearity?

First, we examine the Fe(II)(NH$_3$)$_6$ and Fe(II)(H$_2$O)$_6$ complexes for which we have reference CASPT2 electron densities and magnetizations. Recall that GGA Fe partial charges for the hexa-aqua complex ca. 1.49-1.52 are already in good agreement with CASPT2 (q = 1.52), whereas the GGA partial charges (ca. 1.37 on Fe) for the hexa-ammine complex are in poor agreement (CASPT2 q = 1.62). Considering also magnetic moments reveals that the partial charges alone do not guarantee good agreement with the WFT reference, as GGA hexa-aqua Fe magnetic moments (3.66-3.72 $\mu_B$) underestimate the CASPT2 reference (3.88 $\mu_B$). The hexa-ammine magnetic moments similarly underestimate (3.64-3.72 $\mu_B$) the CASPT2 reference (3.87 $\mu_B$). The underestimation of magnetic moments may be rationalized as a consequence of SIE in mid-row transition metal complexes. Namely, both low-energy majority- and minority-spin states participate in bonding with ligand states, but these are the solely occupied states for the minority-spin in HS complexes. When approximate SIE corrections localize electron density to the ligands, this affects a higher proportion of minority-spin molecular states, and all three tuning procedures thus increase the magnetic moment (Supporting Information Table S6). Sensitivities of magnetic moments are comparable to the sensitivities of partial charges for each method with about 0.33-0.34 $\mu_B$/HFX for global HF exchange, 0.24 $\mu_B$/10 eV of $U$ in DFT+U, and 0.25-0.26 $\mu_B$/bohr$^{-1}$ in range-separated hybrid tuning for both complexes.

We now identify the extent to which DFT Fe partial charges and magnetic moments may be simultaneously matched to CASPT2 values, where the latter are chosen to represent the correct SIE-free electronic structure properties. For each tuning approach, we compute the root-sum-squared (RSS) error of these two quantities for an arbitrary tuning parameter, $i$:

$$\text{RSS}_i = \sqrt{[q(\text{Fe})_i - q(\text{Fe})_{\text{CASPT2}}]^2 + [m(\text{Fe})_i - m(\text{Fe})_{\text{CASPT2}}]^2} \qquad (15)$$

where the first term represents the Fe partial charge error and the second term represents the Fe magnetic moment error. It is useful to assess the RSS error at several points: i) GGA with either the PWBS or LBS, ii) the point of minimum RSS error for each tuning approach, and iii) the point at which LUMO error has been eliminated via the relevant tuning approach. For the two cases considered, GGA RSS errors are generally maxima, and all tuning strategies reduce the RSS errors (Figure 10). The overall minima are near zero for global- or range-separated hybrid tuning on the hexa-ammine complex, albeit at values not typically employed in electronic structure calculations of around 65% or 0.62 bohr$^{-1}$, respectively, whereas DFT+U RSS errors do not approach zero within a range of reasonable $U$ values (i.e., < 10 eV, see Figure 10). DFT+U diverges slightly from the other two approaches by modifying the charge density more slowly.

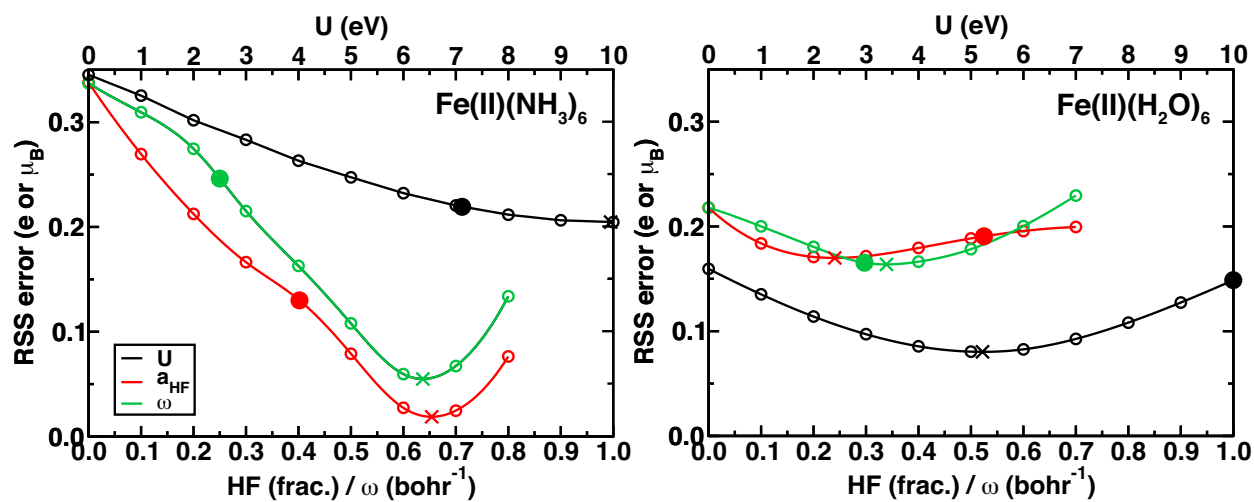

**Figure 10.** RSS partial charge and magnetization error for Fe(II)(NH$_3$)$_6$ (left) and Fe(II)(H$_2$O)$_6$ (right) as a function of tuning parameter. Results for DFT+U (black line), global HF exchange (red line), and range-separated hybrids (green line) are shown with axes corresponding to units for the DFT+U shown at top and hybrid functionals at bottom. All computed points are shown as open circles with a cubic spline, the point that corresponds to zero LUMO error is shown as a large filled circle, and the point that is the minimum on each curve is shown as an X.

Conversely, already good GGA Fe partial charges that worsen with functional tuning in the hexa-aqua complex mean that although magnetic moments are improved, overall RSS errors are lower for DFT+U because it worsens partial charges at a slow rate as it improves magnetic moments, for an optimal agreement near a $U$ of 5 eV. Finally, we consider whether the point of minimum density error coincides with the point of LUMO error elimination for the three methods. Overall, we find that range-separation tuning is the most efficient at eliminating LUMO energetic errors but comparably efficient to global hybrid tuning at eliminating density errors. Thus, for cases such as the hexa-ammine complex where GGA hybridization density errors are significant, the curvature-corrected point leaves behind the most density error for range-separated hybrid tuning followed by DFT+U and global-hybrids. In the hexa-aqua case, RSS errors are closer to their minima values at energy-correction points but mainly due to a balance of increasing errors in partial charges with still-reducing errors in magnetic moments. These two cases are far from all-encompassing but highlight the likelihood that density-based delocalization error corrections may not coincide with those that eliminate energy errors, which may be relevant for prediction of magnetic moments or other observables that depend upon the ground state electron density.

We now compare the equivalence of the three tuning approaches on both LUMO error corrections and density changes over a 15-molecule subset of the original 32 complexes, which includes the full group 16 series, ammonia, phosphine, carbonyl, chloride iron complexes as well as the ammonia, water, and carbonyl titanium complexes (Supporting Information Tables S7-S8). The largest GGA LUMO errors are observed for small, weak-field ligands (e.g., Fe(II)(H$_2$O)$_6$), consistent with previous work[56], and the smallest are observed for electron-rich molecules (e.g., Fe(II)(H$_2$Te)$_6$). For equilibrium complexes, higher LUMO error generally led to higher $p^*$ values for LUMO error elimination, e.g. $U$ of 10.5 eV, HF exchange of 52% or ω of 0.3 bohr$^{-1}$ for Fe(II)(H$_2$O)$_6$ versus considerably smaller values of $U$ of 4 eV, 32% HF exchange, or ω of 0.25 bohr$^{-1}$ for Fe(II)(H$_2$Te)$_6$. The range of tuning parameters to eliminate LUMO errors across complexes for each method, $U^*$ from 4 eV to 15 eV, $a_{HF}^*$ from 0.29 to 0.77, and $\omega^*$ from 0.13 to 0.36 bohr$^{-1}$ may be used to establish ratios for tuning parameters (Supporting Information Figure S5). Moderate correlations between $p^*$ values are observed ($R^2$=0.34 between DFT+U and global hybrids, $R^2$=0.42 between $\omega$-tuning and global hybrids, See Supporting Information Figure S5). These comparisons suggest that DFT+U values near around 5 eV correlate well to the effect of the widely employed 20-25% HF exchange range, and that generally a range of around 12 eV of $U$ correlates well to tuning HF exchange from 0 to 100%. Conversely, the small ranges over which $\omega$-tuning eliminates LUMO errors suggests that increasing $\omega$ to 0.33 bohr$^{-1}$ is equivalent to 100% global HF exchange for the systems considered (see Supporting Information Figure S5). Thus, comparison of rates of piecewise-linearity recovery provide reasonable values relating parameters employed in one tuning approach (e.g., DFT+U) to another (e.g., global hybrid tuning).

Having considered the rate at which each method eliminates LUMO errors, we now may compare how much each method alters the electronic structure by evaluating the metal (M) electron loss at the point of curvature elimination ($\Delta q_M^*$). We obtain $\Delta q_M^*$ by multiplying the linearly-approximated sensitivity with the parameter that eliminates LUMO error:

$$\Delta q_M^*(p) = S_p p^* , \qquad (16)$$

and compare the range of values obtained across methods and complexes. A weak correlation between $p^*$ and $S_p$ leads to greater correlation of $\Delta q_M^*$ between tuning procedures than was obtained for $p^*$ ($R^2$=0.76-0.90, see Supporting Information Figure S6). An examination of the obtained correlations reveals a 0.9:1.0 ratio for charge-loss from DFT+U vs. global HF exchange and 0.5:1.0 for $\omega$-tuning vs. HF exchange. The largest charge losses with global exchange are observed for stretched iron hexa-aqua and hexa-carbonyl complexes at around 0.36 e and 0.24 e, respectively (Supporting Information Table S9).

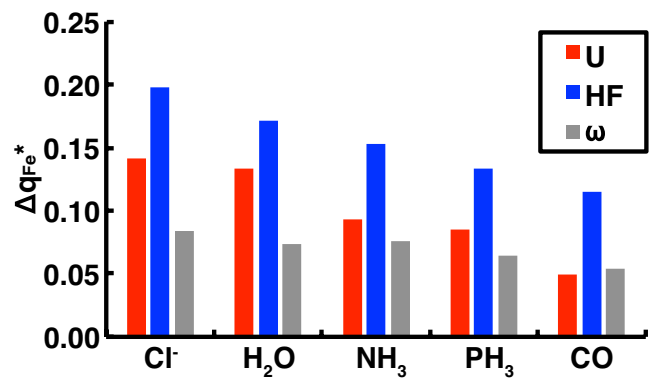

**Figure 11**. Comparison of Fe partial charge loss at point of LUMO error elimination ($\Delta q_{Fe}^*$ in e) for $U$ (red bars), HF exchange (blue bars), and ω (gray bars)-tuning in representative octahedral Fe(II) complexes in order of increasing ligand-field strength from left to right: $Cl^-$, $H_2O$, $NH_3$, $PH_3$, and CO ligands.

We may also interpret the shift in charge density through the nature of the chemical bonding in the complexes. Namely, increasing ligand field strength paradoxically corresponds to *decreased* electron density removal from the metal to the ligand (Figure 11). This observed trend is strongest for global hybrid and DFT+U tuning, with the smaller overall values of charge loss for range-separated hybrids leading to more complex-independent values (see Figure 11 and Supporting Information Table S9). Although some of us have previously identified[56] that LUMO error is generally smaller for strong-field than weak-field ligands, the diminishing $\Delta q_{Fe}^*$ observed here with increasing ligand field strength cannot be explained by this effect alone, as the $p^*$ values are not monotonically decreasing (see Supporting Information Table S6b). It is also not evident that this trend is consistent with the extent of density delocalization error observed in these complexes. Beyond ligand field theory alone, stronger ligands correspond to increased covalency in the M-L bond, which is overestimated by semi-local DFT functionals, as exemplified by more neutral partial charges for the hexa-ammine complex than hexa-aqua, bringing the former into worse agreement with CASPT2 references. Thus, these results suggest that density-derived errors in energy-based functional tuning may over-localize the density for weak M-L bonds but underlocalize the density for strong M-L bonds. This further motivates the development of strategies that separately correct density-derived and energetic self-interaction errors for instance by employing DFT+U for density errors and range-separated hybrids for energy delocalization errors.

## 5. Conclusions

We have compared three diverse strategies for mitigating self-interaction error within approximate DFT, i.e., DFT+U, global hybrid tuning, and range-separated hybrid tuning, and we identified that these three methods have qualitatively equivalent behavior across the 32 transition metal complexes considered in this work. Although SIE is known to increase unphysical electron delocalization, the universal nature of electron localization by SIE-reducing methods from the metal to the ligand had not yet been noted. Indeed, regardless of valence orbital diffuseness (i.e., from $2p$ to $5p$), ligand electronegativity (i.e., from Al to O), basis set (i.e., plane wave versus localized basis set), metal (i.e., Ti, Fe, Ni) and spin state, or tuning method, we consistently observe substantial charge loss at the metal and gain at ligand atoms (ca. 0.3-0.5 e).

We further distinguished energy-derived delocalization error, i.e., deviations from piecewise linearity, from density-derived errors, as observed through comparison of metal-centered partial charges and magnetic moments for representative complexes from approximate DFT versus CASPT2 references. We observed increased density errors with ligand field strength or hybridization but simultaneously decreased overall impact of the tuning methods on the electron density. Generally, the minimum error in partial charges and magnetic moments was observed to occur at higher tuning parameters, particularly for range-separation, than those that eliminated energy delocalization error alone. These observations suggest that multi-faceted error correction approaches that separately treat density delocalization and energetic errors are needed in order to recover both correct density and magnetization properties at integer electrons and molecular orbital energies. The development of such flexible corrections for transition metal chemistry is underway within our group.

ASSOCIATED CONTENT

**Supporting Information**. Coordinates of optimized geometries; list of pseudopotentials used; list of complexes studied; comparison of selected DFT and CASPT2 partial charges and spin densities; metal and ligand projected densities of states for selected complexes; Bader and NPA charges at all values of tuning parameters; sensitivity vs. GGA partial charge; comparison of Bader vs. NPA partial charges and further discussion; LUMO error tuning details and comparison among tuning methods. This material is available free of charge via the Internet at http://pubs.acs.org.

AUTHOR INFORMATION

**Corresponding Author**

*email: hjkulik@mit.edu phone: 617-253-4584

**Notes**

The authors declare no competing financial interest.

ACKNOWLEDGMENT

The authors acknowledge partial support by the National Science Foundation under grant number ECCS-1449291. H.J.K. holds a Career Award at the Scientific Interface from the Burroughs Wellcome Fund. This work was carried out in part using computational resources from the Extreme Science and Engineering Discovery Environment (XSEDE), which is supported by National Science Foundation grant number ACI-1053575. The authors thank Adam H. Steeves for providing a critical reading of the manuscript.

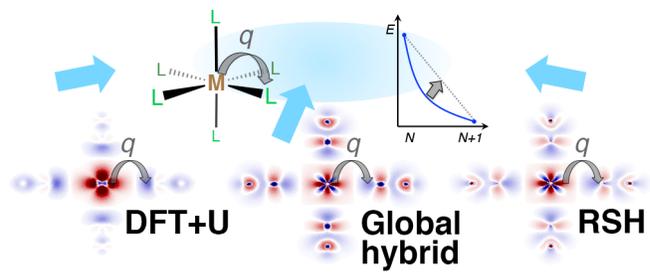